\documentstyle[preprint,prb,aps]{revtex}

\def\AmS{{\protect\the\textfont2
        A\kern-.1667em\lower.5ex\hbox{M}\kern-.125emS}}

\makeatletter
\def\thepage{1-\@arabic\c@page}
\def\@pnumwidth{2em}
\makeatother
\begin{document}
\draft

\title{Pressure Effect on the Spin Fluctuations in YMn$_2$\\
--Mn NQR Study--}

\author{K.~Nishikido$^{\dag}$, C.~Thessieu$^{\dag}$, G-q.~Zheng$^{\dag}$,
Y.~Kitaoka$^{\dag}$, K.~Asayama$^{\dag}$, R.~Hauser$^{\ddag}$}
\address{$^{\dag}$ Dpt of Mat. Physics, Osaka University, Osaka 560}
\address{$^{\ddag}$ Institut f\"ur Experimentalphysik, a-1040 Wien, Austria}
\date{August 25, 1997}
\maketitle

\begin{abstract}
We report the NQR study on the pressure-induced paramagnetic state in the
antiferromagnetic (AF) intermetallic compound YMn$_2$ with the N\'eel
temperature $T_{\text{N}}$=100~K at ambient pressure. From the $T$
variation of the nuclear spin-lattice relaxation rate, $^{55}(1/T_1)$ of $%
^{55}$Mn above the critical pressure of 4~kbar, the spin fluctuation feature
is found to change below a temperature $T_{\text{sf}}$. At higher
temperature than $T_{\text{sf}}$, the nuclear relaxation behavior is well
described in terms of the self-consistent renomalized (SCR) spin fluctuation
theory for nearly AF metals, whereas below $T_{\text{sf}}$ a deviation is
significant and $1/T_1T$ exhibits a weak $T$ variation. It is pointed out
that $T_{\text{sf}}$ coincides with the temperature below which a $T^{2}$
law in electrical resistivity is valid as expected for a Fermi liquid ground
state. We proposed that these features are understood from the standpoint
that the development of AF spin fluctuation remains in short-range
associated with the singlet formation among Mn spins in each tetrahedron as
suggested by the inelastic neutron experiments on Y$_{1-x}$Sc$_x$Mn$_2$.
\end{abstract}

\pacs{ }

\section{Introduction}

In a series of intermetallic compound RMn$_2$ (R=Rare Earth),
an existence of magnetic moments on Mn sites depends largely on an
inter-atomic Mn-Mn distance\cite{HAU94}. Above a critical distance of $d_{\text{%
Mn-Mn}} $=2.7~$10^{-10}$m there exists a magnetic moment on the Mn which is
antiferromagnetically ordered in general. Below this critical distance the
ground state remains paramagnetic down to lowest temperature. This fact that
the Mn-Mn inter-atomic distance plays a key-role in determining their
magnetic properties indicates that these compounds may be magnetically
sensitive to such a perturbation as a chemical substitution or an
application of pressure. An interesting possibility is then that an
application of pressure induces Quantum Critical Phase Transition governed
by the zero point quantum spin fluctuation. In particular this is the case
of YMn$_2$ which exhibits antiferromagnetic (AF) order with a N\'eel
temperature around 100~K at ambient pressure. A relatively low pressure of $%
P_{\text{c}}$=3 to 4~kbar \footnote{%
The value of the critical pressure depends on the experimental technique.}
is enough to suppress the AF order\cite{OOM87}. The paramagnetic to AF magnetic
transition upon cooling is of first order type accompanying by well
pronounced hysteretic phenomena in various measurements. Therefore, the AF
to paramagnetic phase transition across $P_c$ is anticipated to be first
order type. So far it is reported that the paramagnetic state at ambient
pressure is dominated by the AF spin fluctuations \cite{DEP87} which is well treated
in terms of the self-consistent renormalized (SCR) spin fluctuation theory
developed by Moriya et al. \cite{MOR85}. As a matter of fact, the Sc-substitution of Y
suppresses the AF order and the spin fluctuations in the paramagnetic phase
was reported to be well described in terms of the SCR theory \cite{WAD87}. On the
other hand, the inelastic neutron scattering experiment has found that the
singlet correlation among four Mn spins forming the tetrahedron is developed
to stabilize this paramagnetic phase, i.e., a kind of spin liquid state is
formed in the Sc substituted system. Thus, it is still controversial whether
or not this novel character of spin fluctuations is actually related to the
nature in the paramagnetic state above $T_{\text{N}}$ in YMn$_2$ because
the effect of disorder induces by the Sc -substitution is not addressed yet.
In other words, we address the question whether or not, near the critical
point in YMn$_2$ where the N\'eel temperature is suppressed to zero, the AF
spin fluctuations are also critical with a damping energy tending to zero.
An application of pressure to suppress the AF order is promising to settle
these issues, because this is the only way to stabilize the paramagnetic
state in YMn$_2$ at low temperatures without any disorderliness. The aim of
this paper is to enligthten the evolution of low-energy magnetic excitations
in the pressure-induced paramagnetic phase by approaching the critical point
from the higher pressure side.

\section{Experimental Details}

A polycrystalline sample has been crushed in fine powder in
order to ensure the penetration of the radio-frequency pulsed field. A
conventional clamp-type pressure cell was employed for the NMR experiment
under pressure with use of the Fluorinert liquid as the pressure
transmitting medium. The Fluorinert volume was of the proportion as the
sample volume and froze below 200~K. To calibrate the pressure at the sample
position at low temperature, the pressure shift in $T_{\text{c}}$ of lead
powder was monitored. The measurement of $T_{\text{c}}$ was performed by
an inductive method using the in-situ NMR coil in a resonant circuit. The
NQR spectrum has been obtained by plotting the spin-echo intensity as a
function of frequency with an interval of 50~kHz. The spin-lattice
relaxation time, $T_1$, was measured by the saturation recovery method in a
temperature range of $4.2-120$~K. For the NQR transition of $\pm
5/2\leftrightarrow \pm 3/2$, the recovery function of nuclear magnetization,
M, is given by
\begin{equation}
\frac{M(\infty )-M(t)}{M(\infty )}=\frac 37\exp (-\frac{3t}{T_1})+\frac
47\exp (-\frac{10t}{T_1})
\label{equ1}
\end{equation}
and the spin-lattice relaxation time was uniquely determined by a fit of the
data with the above formula.

\section{Results and Analysis}

Fig.\ \ref{fig1} indicates the evolution of the $^{55}$Mn NQR spectrum
for the $\pm 5/2\leftrightarrow \pm 3/2$ transition at $4.2$~K under pressure.
With increasing pressure, the spectrum experiences a slight shift towards a
higher frequency side, reflecting an increase of the electric field gradient
(EFG). The full-width at half-maximum in a series of NQR spectra does not
change appreciably, which means that the pressure-induced distribution in
EFG can be ignored.
The nuclear spin-lattice relaxation time $T_1$ was not
measured above 120~K in order to avoid a melting of the transmitting liquid
medium at high temperatures which increases the pressure effectively . The $%
T $ dependence of $1/T_1T$, which is plotted with the semi-logarithmic scale
in temperature (Fig.\ \ref{fig2}), shows the possible existence of two regimes
marked by a kink (arrows). Hereafter, we will refer to this cross-over
temperature as a characteristic spin fluctuation temperature, $T_{\text{sf}%
}$, in the pressure-induced paramagnetic phase. $T_{\text{sf}}$ increases
from 30~K with increasing pressure. Below $T_{\text{sf}}$, $1/T_1T$
becomes weakly temperature dependent.
The spin-lattice relaxation rate, $1/T_1$, is in
general related to the transverse component of the imaginary part in the
dynamical magnetic susceptibility $Im\chi ^{\bot }(q,\omega )$ such as 
\begin{equation}
1/T_1=\gamma _{\text{N}}^2N_0^{-2}T\sum_qA^2(q) \frac{Im\chi
^{\bot}(q,\omega _0)}{\omega _0}
\label{equ2}
\end{equation}
where $\gamma _{\text{N}}$ is the gyromagnetic ratio of the nuclei
studied, $N$ the number of atoms, $A(q)$ the hyperfine coupling constant, and $%
\omega _0$ the NMR frequency ($\bot$ represents the transverse component of
the generalized susceptibility). In the case of nearly antiferromagnetic
metals, the q dependence of the low-energy spin fluctuations are dominated
by the components close of the AF vector \textbf{q}=\textbf{Q} and the main
contribution will thus be such as $\chi ^{\bot }(\mathbf{Q},\omega _0,T)$.
Therefore Eq.\ (\ref{equ2}) can be given by a simple form of 
\begin{equation}
1/T_1T\propto A^2\times Im\chi ^{\bot }(\mathbf{Q},\omega _0,T).
\label{equ3}
\end{equation}
The pressure and thermal dependences of $1/T_1T$ thus allow us to probe
corresponding dependences of AF spin fluctuation. In the framework of the
SCR theory the dynamical susceptibility is given by 
\begin{equation}
\chi ^{\bot }(q, \omega )=\frac{\chi _0^{\bot }(q, \omega )}{1-U\chi
_0^{\bot }(q, \omega )+\lambda (q, \omega)}
\label{equ4}
\end{equation}
where $\lambda (q, \omega)$ is the parameter to make the self-consistency
equation valid, the non-interacting susceptibility $\chi _0^{\bot }(q,
\omega )$ and $U $ the Coulomb intra-site interaction. The denominator of
Eq.\ (\ref{equ4}) is equivalent to the Stoner factor which tends to zero at the
criticality and $P_{\text{c}}$. Correspondingly the divergence of $%
Im\chi^{\bot}(q,\omega)$ means the divergence in the static staggered
susceptibility and as a result, it is expected that $1/T_1T $ undergoes a
divergence at $P_{\text{c}}$ or it is largely enhanced approaching $P_{%
\text{c}}$. However, the experiment does not follow the above expectation.
Rather, the development of AF spin fluctuations is saturated upon cooling
and approaching the $P_{\text{c}}$. Namely, $T_{\text{sf}}$ does not
tend to zero, but seems to remain finite even at $P_{\text{c}}$. This may
due to the first order nature of the transition from the AF state to
paramagnetic at $P_{\text{c}}$ induced by pressure. Above $T_{\text{sf}}$%
, the spin fluctuations are dominated by thermal fluctuations which are
treated in terms of the SCR theory, whereas 1/$T_1T$ approaches a constant
value below $T_{\text{sf}}$ characteristic for the Fermi liquid excitation.
Fig.\ \ref{fig3} indicates the pressure dependence of $T_{\text{sf}%
} $ deduced from the present $T_1$ measurement. From the electrical
resistivity measurements under pressure, the spin fluctuation temperature
was defined as the cross-over temperature below which the thermal variation
of $\rho(T)$ obeys the expected Fermi liquid $T^2$ law\cite{HAU94}. Remarkably, a
fairly good agreement is obtained concerning the variation of both $T_{%
\text{sf}}$ under pressure 
\begin{equation}
\frac{\partial \ln T_{\text{sf}}}{\partial P}=59 \text{{Mbar}}^{-1}
\label{equ5}
\end{equation}
With these experimental implications, it is possible that the development in
spin fluctuation remains in short-range upon cooling.

\section{Conclusion}

From the thermal variation of the nuclear spin lattice
relaxation rate, $^{55}(1/T_1T)$ of $^{55}$Mn above the critical pressure of
4~kbar, the spin fluctuation feature has been found to change below the
temperature $T_{\text{sf}}$. At temperatures higher than $T_{\text{sf}}$%
, the nuclear relaxation behavior is well described in terms of the
self-consistent renormalized spin fluctuation theory for a nearly AF metal,
whereas below $T_{\text{sf}}$, a deviation from that is significant and $%
1/T_1T$ exhibits a weak temperature variation. It is pointed out that $T_{%
\text{sf}}$ coincides with the temperature below which a $T^2$ law in
resistivity is valid as expected for the Fermi liquid state. We proposed
that these features are understood from the view that the development in AF
spin fluctuation remains in short-range associated with the singlet
formation among Mn spins in each tetrahedron as suggested by the inelastic
neutron experiments on Y$_{1-x}$Sc$_x$Mn$_2$.

\begin{figure}
\caption{Pressure effect on the electric quadrupole frequency of
the $^{55}$Mn resonance $\pm 5/2 \leftrightarrow \pm 3/2$ at $T=4.2$~K.}
\label{fig1}
\end{figure}

\begin{figure}
\caption{Pressure effect on the thermal dependence of 1/$T_1T$
of the $^{55}$Mn nuclei in YMn$_2$.}
\label{fig2}
\end{figure}

\begin{figure}
\caption{Comparaison of the pressure effect on the spin fluctuations temperature,
$T_{\text{sf}}$, in YMn$_2$ defined upon the present NQR experiment
($\bullet$)
and by previous authors upon electrical resistivity measurements ($\circ$).}
\label{fig3}
\end{figure}

\end{document}